\documentstyle[11pt,aaspp4,flushrt]{article} 
\input psfig
\begin{document}
\title{Flares on the Black Holes} 
\author{Andrei Gruzinov}
\affil{Institute for Advanced Study, School of Natural Sciences, Princeton, NJ 08540}

\begin{abstract}
A Kerr black hole can flare quasi-periodically, if it is magnetically connected to the accretion disk. The flares extract the rotational energy of the hole. The flaring occurs because the hole twists the magnetic field lines beyond the threshold of the screw instability. 

\end{abstract}

\section{Introduction}
We show that a Kerr black hole can flare quasi-periodically, if it is connected to the accretion disk by a magnetic field. The averaged luminosity in the flares is $\sim B^2r_+^2c$, where $B$ is the magnetic field at the event horizon, $r_+$ is the event horizon radius, $c$ is the speed of light. 

The mechanism considered here is different from the Blandford-Znajek process - the magnetic field connects the hole to the accretion disk, instead of to infinity. In \S 2 we give the astrophysical formulation of the problem and an order-of-magnitude solution. In \S 3 we idealize the problem and solve it rigorously.

\section{Astrophysics of Black Hole Flares}
Fig. 1. shows a Kerr black hole, its accretion disk, and a magnetic spot in the disk. The hole will twist the magnetic field lines, that is a toroidal magnetic field will be generated. To calculate the toroidal field, we treat the hole as a rotating conducting ball, radius $\sim r_+$, conductivity $\sigma \sim c/r_+$ (Znajek 1978). We assume for simplicity that the ball is rotating at about the speed of light, corresponding to a nearly maximally rotating black hole. In the ball rest frame, the poloidal electric field is $E\sim B$, where $B$ is the magnetic field. The current is $I\sim r_+^2\sigma E\sim cr_+B$. This current creates a toroidal magnetic field, $B_t\sim I/(cr_+)\sim B$. 

\begin{figure}[htb]
\psfig{figure=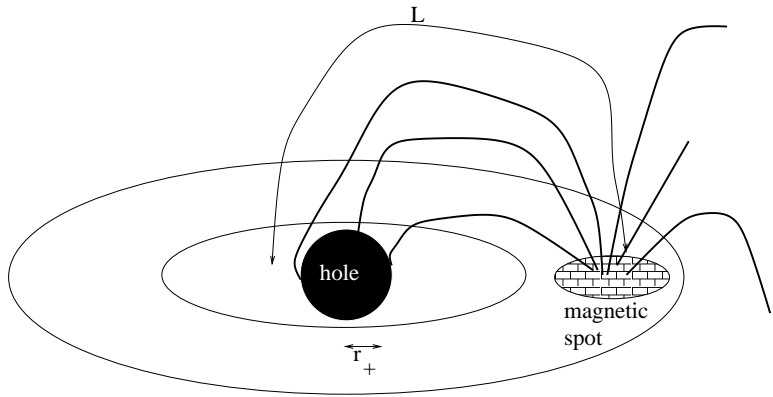,width=4in}
\caption{Kerr black hole in a magnetic contact with the accretion disk.}
\end{figure}

The twisted magnetic field will try to unwind itself. According to the Kruskal-Shafranov criterion (Kadomtsev 1966), this leads to the screw instability when the toroidal field becomes so strong that the magnetic field line turns around itself about once. The instability condition is $B_t/r_+\gtrsim B/L$, or $L\gtrsim r_+$, where L is the distance from the event horizon to the magnetic spot measured along the field line. 

If the stationary magnetic configuration is screw-unstable, the black hole magnetosphere flares and releases part of the magnetic energy. Then the hole twists the magnetic field again, and the magnetosphere becomes screw-unstable. The flares repeat quasi-periodically, with the period $\sim r_+/c$.

\section{Idealized problem}
We straighten the field lines connecting the hole and the magnetic spot in the disk. We neglect the motion of the spot as compared to the hole rotation. This transforms Fig.1 into Fig.2.

Consider a Kerr black hole and two ideally conducting plates at a distance $L$ from the hole. The plates are perpendicular to the black hole angular momentum. There is a uniform external magnetic field $B_0$ frozen into the plates. The electromagnetic field is force-free. We calculate the force-free magnetosphere in \S 3.1, formulate the stability problem in \S 3.2, and show that the magnetosphere can be screw-unstable in \S 3.3. 

The results given here constitute a rigorous proof that the magnetosphere of a Kerr black hole is screw-unstable for certain values of $L\gg r_+\gg a$, where $a$ is the angular momentum of the hole per unit mass. In fact, this means that a fast, $a\sim r_+$, Kerr hole will flare if $L>Cr_+$ where $C$ is a dimensionless number of order unity.

\begin{figure}[htb]
\psfig{figure=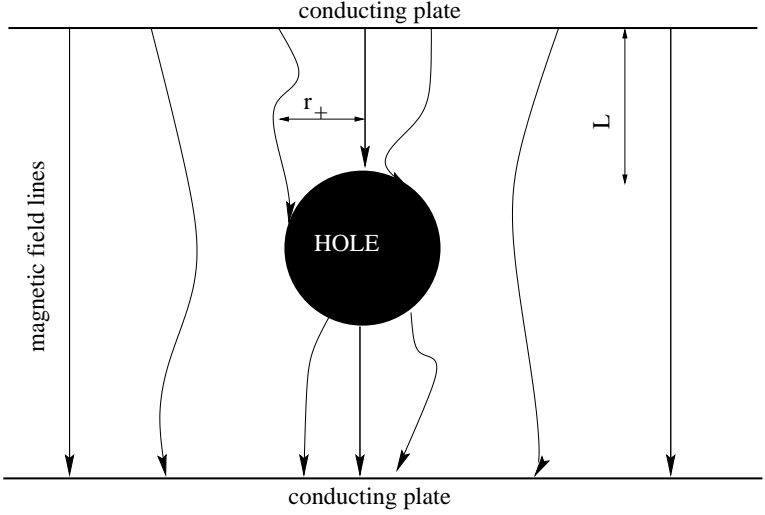,width=4in}
\caption{Kerr black hole in a magnetic field. At a distance $L$ from the black hole, the magnetic field is frozen into ideally conducting plates.}
\end{figure}

\subsection{The force-free magnetosphere}
We derive the equation for the force-free magnetosphere following the steps similar to Blandford \& Znajek (1977). In our case the equations and their derivation are much simpler because the electrostatic potential is everywhere zero due the ideally-conducting plates.  

The Boyer-Lindquist coordinates are used, with $c=1$, and $G=1$. The metric is
\begin{equation}
ds^2=(1-{2Mr\over \Sigma })dt^2+{4Mar\over \Sigma }\sin ^2\theta dtd\phi -{\Sigma \over \Delta }dr^2-\Sigma d\theta ^2-{(r^2+a^2)^2-\Delta a^2\sin ^2\theta \over \Sigma }\sin ^2\theta d\phi ^2,
\end{equation}
\begin{equation}
\Sigma =r^2+a^2\cos ^2\theta ,
\end{equation}
\begin{equation}
\Delta =r^2-2Mr+a^2,
\end{equation}
$M$ is the mass, $a$ is the angular momentum per unit mass. The event horizon radius is $r_+=M+\sqrt{M^2-a^2}$.

The magnetosphere is described by the Maxwell equations
\begin{equation}
\partial _k(\sqrt{-g} F^{ik})=-4\pi \sqrt{-g} j^i,
\end{equation}
and the force-free condition
\begin{equation}
F_{ik}j^k=0.
\end{equation}
Here $g={\rm det}(g_{ik})=-\Sigma ^2\sin ^2\theta$, $F_{ik}=\partial _iA_k-\partial _kA_i$ is the electromagnetic field tensor, $A^i$ is the four-potential, $j^i$ is the four-current. We consider the stationary axisymmetric case, $\partial_t=\partial _{\phi }=0$.

From (5),
\begin{equation}
(j^r\partial _r+\j^{\theta }\partial _{\theta })A_t=0,
\end{equation}
\begin{equation}
(j^r\partial _r+\j^{\theta }\partial _{\theta })A_{\phi }=0.
\end{equation}
Then $A_t$ must be a function of $A_{\phi }$. At the plates $A_t=0$. Assuming that all the points in the magnetosphere are connected to the conducting plates by the magnetic field lines,  $A_t=0$ everywhere. 

Denote $A\equiv -A_{\phi }$, $B\equiv -\partial _rA_{\theta }+\partial _{\theta }A_r$. Then the remaining two force-free conditions and three of the Maxwell equations \footnote{The remaining Maxwell equation gives the charge density. We do not need it.} are
\begin{equation}
j^{\phi }\partial _rA+Bj^{\theta }=0,
\end{equation}
\begin{equation}
j^{\phi }\partial _{\theta }A-Bj^r=0,
\end{equation}
\begin{equation}
\partial _{\theta }(\sqrt{-g}g^{rr}g^{\theta \theta }B)=4\pi \sqrt{-g}j^{r},
\end{equation}
\begin{equation}
\partial _r(\sqrt{-g}g^{rr}g^{\theta \theta }B)=-4\pi \sqrt{-g}j^{\theta },
\end{equation}
\begin{equation}
\partial _r(\sqrt{-g}g^{\phi \phi }g^{rr}\partial _rA)+\partial _{\theta }(\sqrt{-g}g^{\phi \phi }g^{\theta \theta}\partial _{\theta }A)=-4\pi \sqrt{-g}j^{\phi }.
\end{equation}
The combination $\sqrt{-g}g^{rr}g^{\theta \theta }B=2I$, where $I$ is the poloidal current. From (10) and (11),
\begin{equation}
(j^r\partial _r+\j^{\theta }\partial _{\theta })I=0.
\end{equation}
From (7) and (13), $I=I(A)$ - the current flows along the field lines. From (10), $2 \pi \sqrt{-g} j^{r}=I'\partial _{\theta }A$, where $I'\equiv dI/dA$. Now from (9), 
\begin{equation}
j^{\phi }=-{II'\over \pi g g^{rr}g^{\theta \theta }}.
\end{equation}

With $j^{\phi }$ given by (14), $j^r$ by (10), and $j^{\theta }$ by (11), equations (7)-(11) are satisfied. Equation (12) takes the final form
\begin{equation}
\partial _r(g_{00}\partial _rA)+\Delta ^{-1}\sin \theta \partial _{\theta }( (g_{00}/\sin \theta )\partial _{\theta }A)+4\Delta ^{-1}\Sigma II'=0,
\end{equation}
where $g_{00}=1-2Mr\Sigma ^{-1}$. The boundary conditions at the plates are
\begin{equation}
A|_{z=\pm L}=B_0\rho ^2/2,
\end{equation}
where $z\equiv r\cos \theta$, $\rho \equiv r\sin \theta$. The boundary condition at the event horizon follows from the requirement that $F^2$ be finite at $r_+$. This gives 
\begin{equation}
2I={a\sin \theta \over \Sigma}\partial _{\theta }A|_{r=r_+}.
\end{equation}
We now solve (15) assuming $a\ll r_+$. For $a=0$, (17) gives $I=0$, and then (15) and (16) give $A=B_0r^2\sin ^2\theta /2$. We use this $A$ in (17) and find the poloidal current,
\begin{equation}
I(A)={aA\over r_+^2}\sqrt{1-{2A\over B_0r_+^2}}.
\end{equation} 
We can now use this $I(A)$ in (15) and iterate if greater precision is required. But we do not need a more accurate solution for our current purpose. Equation (18) is sufficiently accurate to analyze the stability of the magnetosphere in the limit of $r_+\gg a$.

\subsection{Stability in Force-Free Electrodynamics}
We have found the stationary magnetic configuration. Now we study its stability in the framework of force-free electrodynamics (FFE) defined by (4), (5). For $L\gg r_+$, we can use the galilean metric. 

FFE is classical electrodynamics supplemented by the force-free condition. Thus
\begin{equation}
\partial _t{\bf B}=-\nabla \times {\bf E},
\end{equation}
\begin{equation}
\partial _t{\bf E}=\nabla \times {\bf B}-{\bf j},
\end{equation}
\begin{equation}
\rho {\bf E}+{\bf j}\times {\bf B}=0.
\end{equation}
$\nabla \cdot {\bf B}=0$ is the initial condition. The speed of light is $c=1$; $\rho =\nabla \cdot {\bf E}$ and ${\bf j}$ are the charge and current densities multiplied by $4\pi$. The electric field is everywhere perpendicular to the magnetic field, ${\bf E}\cdot {\bf B}=0$ \footnote{If $\rho \not= 0$, ${\bf E}\cdot {\bf B}=0$ follows from (21), if $\rho =0$, this condition is an independent basic equation of FFE.}. The electric field component parallel to the magnetic field should vanish because the charges are freely available in FFE. It is also assumed that the electric field is everywhere weaker than the magnetic field, $E^2<B^2$. Then equation (21) means that it is always possible to find a local reference frame where the field is a pure magnetic field, and the current is flowing along this field. FFE is Lorentz invariant. 

Equation (21) can be written in the form of the non-linear Ohm's law
\begin{equation}
{\bf j}={({\bf B}\cdot \nabla \times {\bf B}-{\bf E}\cdot \nabla \times {\bf E}){\bf B}+(\nabla \cdot {\bf E}){\bf E}\times {\bf B} \over B^2}.
\end{equation}
Equations (19), (20), (22) form an evolutionary system (initial conditions $\nabla \cdot {\bf B}=0$ and ${\bf E}\cdot {\bf B}=0$ are implied). It therefore makes sense to study stability of equilibrium electromagnetic fields in FFE. One can also study linear waves and their nonlinear interactions in the framework of FFE (Thompson \& Blaes 1998).

It is convenient to introduce a formulation of FFE similar to magnetohydrodynamics (MHD); then we can use the familiar techniques of MHD to test stability of magnetic configurations. To this end, define a field ${\bf v}={\bf E}\times {\bf B}/B^2$, which is similar to velocity in MHD. Then ${\bf E}=-{\bf v}\times {\bf B}$ and equation (19) becomes the ``frozen-in'' law
\begin{equation}
\partial _t{\bf B}=\nabla \times ({\bf v}\times {\bf B}).
\end{equation}
From ${\bf v}={\bf E}\times {\bf B}/B^2$, and from equations (19)-(21), one obtains the momentum equation
\begin{equation}
\partial _t(B^2{\bf v})=\nabla \times {\bf B}\times {\bf B}+\nabla \times {\bf E}\times {\bf E}+(\nabla \cdot {\bf E}){\bf E},
\end{equation}
where ${\bf E}=-{\bf v}\times {\bf B}$. Equations (23), (24) are the usual MHD equations except that the density is equal $B^2$ and there are order $v^2$ corrections in the momentum equation.

A force-free magnetic field ($\nabla \times {\bf B}\times {\bf B}=0$) with a zero electric field is a stationary FFE solution. We will study the stability of the force-free magnetic configurations using the MHD formulation of FFE. Linear perturbations, denoted ${\bf b}$ and ${\bf v}$, $\propto \exp (-i\omega t)$, satisfy
\begin{equation}
-i\omega {\bf b}=\nabla \times ({\bf v}\times {\bf B}),
\end{equation}
\begin{equation}
-i\omega B^2{\bf v}=\nabla \times {\bf B}\times {\bf b}+\nabla \times {\bf b}\times {\bf B}.
\end{equation}
Define the displacement \mbox{\boldmath $\xi$} by ${\bf v}=-i\omega \mbox{\boldmath $\xi$}$. From (25),  ${\bf b}=\nabla \times (\mbox{\boldmath $\xi$}\times {\bf B})$. Now equation (26) can be written as
\begin{equation}
-\omega ^2B^2\mbox{\boldmath $\xi$}=\nabla \times {\bf B}\times \nabla \times (\mbox{\boldmath $\xi$}\times {\bf B})+\nabla \times \nabla \times (\mbox{\boldmath $\xi$}\times {\bf B})\times {\bf B}\equiv -\hat{\bf K} \mbox{\boldmath $\xi$}.
\end{equation}

Since the operator $\hat{\bf K}$ is self-adjoint (Kadomtsev 1966 and references therein), the frequency is given by the variational principle
\begin{equation}
\omega ^2={\rm min}{ \int d^3r\mbox{\boldmath $\xi$}\hat{\bf K}\mbox{\boldmath $\xi$} \over \int d^3rB^2\mbox{\boldmath $\xi$}^2 }.
\end{equation}
The equilibrium field ${\bf B}$ is unstable if the potential energy
\begin{equation}
W\equiv \int d^3r\mbox{\boldmath $\xi$}\hat{\bf K}\mbox{\boldmath $\xi$},
\end{equation}
is negative for some displacements \mbox{\boldmath $\xi$}. 

Consider the case of cylindrical symmetry, coordinates $(r,\theta ,z)$. The equilibrium field ${\bf B}=(0,U(r),B(r))$ should satisfy $BB'+Ur^{-1}(rU)'=0$, where the prime denotes the r-derivative. For an eigenmode $\propto \exp (im\theta +ikz)$, the potential energy reduces to (Kadomtsev 1966)
\begin{equation}
W\propto \int dr(f\xi '^2+g\xi ^2),
\end{equation}
where $\xi $ is the radial component of the displacement. The other two components of the displacement vector, $\xi _{\theta }$ and $\xi _z$, were chosen to minimize the energy for a given $\xi $. The functions of radius $f$ and $g$ are given by
\begin{equation}
f=r{(krB+mU)^2\over k^2r^2+m^2},
\end{equation}
\begin{equation}
g=r^{-1}\left( {k^2r^2+m^2-1\over k^2r^2+m^2}(krB+mU)^2+{2k^2r^2\over (k^2r^2+m^2)^2}(k^2r^2B^2-m^2U^2)\right).
\end{equation}

We now show that magnetic configurations with a non-zero toroidal field $U$ are screw unstable. Screw means, e. g., that $m=1$, but not $m=-1$, is unstable for a given sign of $k$. Kadomtsev (1966) gives a clear discussion of the screw instability in plasmas, and shows that the screw mode is the most dangerous mode (if the plasma is unstable, it is screw unstable).    

Assume that $U(0)=0$, $U'(0)>0$, $U\rightarrow 0$ for $r\rightarrow \infty$, and $U$ is positive in between. Assume that $B$ is everywhere positive. Let $k$ be positive and small. Take $m=-1$. Let $r_0$ be the first zero of $krB-U$. Take a trial function $\xi=1$ for $r<r_0$, and zero at $r>r_0$. We can chose this generalized function $\xi $ in such a way that the first term in the energy integral (31) vanishes. The second term is an integral of $g$ from $0$ to $r_0$. For small $k$, $g=k^2r(krB-U)(3krB+U)$ is negative. 

Thus an eigenmode $\propto \exp (-i\theta +ikz)$ is screw-unstable if $krB\lesssim U$. If the length of the magnetosphere in the z-direction is $L$, the minimal value of $k$ is $\sim L^{-1}$. The instability condition (the Kruskal-Shafranov criterion) is 
\begin{equation}
{U\over r}>C{B\over L},
\end{equation}
where $C$ is a dimensionless number of order unity. The value of $C$ depends on the boundary conditions.

\subsection{Stability of the Black Hole Magnetosphere}
For the black hole magnetosphere, in the limit $r_+\gg a$, at a distance from the hole $\gg r_+$, the poloidal field $B$ is approximately constant and equal to $B_0$. The poloidal current is given by (18), and the toroidal field is 
\begin{equation}
U={a\over r_+}B_0{r\over r_+}\sqrt{1-{r^2\over r_+^2}},
\end{equation}
in the cylindrical coordinates of \S 3.2. According to the Kruskal-Shafranov criterion, the magnetosphere will be screw unstable if $L>Cr_+^2/a$, where $C$ is a dimensionless number of order unity.

\section{Discussion}
The flaring is a new mechanism for the extraction of the rotational energy of Kerr holes. Many previous studies considered the Blandford-Znajek mechanism in which the magnetic field connects the event horizon to infinity. By contracts, the flaring occurs when the magnetic field ends in the accretion disk. Both mechanisms give the same averaged luminosity, as expected from dimensional analysis. 

The essential feature of our mechanism is that the rotational frequency of the hole determines the time structure of the radiated energy. The hole frequency may appear observationally as a quasi-periodicity of the X-ray flux. It would be interesting to search for this periodicity in the high-time-resolution data.

\acknowledgements I thank John Bahcall for discussions. This work was supported by NSF PHY-9513835.

\end{document}